**Kinetics and Parameters of Epitaxial Monolayered Continuous large area Molybdenum disulfide Growth**

Rakesh K. Prasad[1], Dilip K. Singh[1*]

[1] Birla Institute of Technology Mesra, Ranchi -835215 (India)

[*]Email: dilipsinghnano1@gmail.com

**Abstract**

The growth of large crystallite continuous monolayer materials like molybdenum disulfide ($MoS_2$) with desired morphology via chemical vapor deposition (CVD) remains a challenge. In CVD, the complex interplay of various factors like growth temperature, precursors, and nature of the substrate decides the crystallinity, crystallite size, and coverage area of the grown $MoS_2$ monolayer. In the present work, we report about the role of weight fraction of molybdenum trioxide ($MoO_3$), Sulfur, and carrier gas flow rate towards nucleation and monolayer growth. The concentration of $MoO_3$ weight fraction has been found to govern the self-seeding process and decides the density of nucleation sites affecting morphology and coverage area. Carrier gas flow of 100 sccm argon results into large crystallite continuous films with lower coverage area (70 %), while the flow rate of 150 sccm results into 92 % coverage area with reduced crystallite size. Through systematic variation of experimental parameters, we have established the recipe for the growth of large crystallite atomically thin $MoS_2$ suitable for optoelectronic devices.

## Introduction

For past nearly a decade, twoï dimensional (2D) semiconducting transition metal dichalcogenides (TMDs), $MX_2$ (M=Mo, W; X=S, Se) have attracted huge attention due to their distinct structural, physical, and chemical properties arising from the layered structure with Van der Waal forces of interaction in between the layers opening up new possibilities in the form of efficient and thin electronic and optoelectronic devices. [1-3]. Molybdenum disulfide ($MoS_2$) is a layered 2D semiconducting material with bandgap in the range 1.2 eV to 1.9 eV, whose optical and electrical properties are dependent on the number of layers [4]. Monolayered $MoS_2$ is an atomically thin directï bandgap semiconductor with considerable enhancement in the photoluminescence, while increase in the number of layers shows indirect bandgap semiconducting behavior [5]. Monolayered $MoS_2$ have high potential application in the domain of optoelectronics, nano-electronics and photonics[1,4].

A number of techniques have been attempted to achieve large crystals of single-layered $MoS_2$. Initially, various types of exfoliation techniques were explored like scotch tape-assisted micromechanical exfoliation [6], liquid-phase exfoliation [7], electrochemical Li-intercalation from bulk natural crystals [8]. These methods allowed to achieve nanometer to few micrometer sizes of monolayered $MoS_2$. To grow films at desired locations with the predefined number of layers having large coverage area, different growth techniques were attempted like thermolysis of ammonium thiomolybdate [9], physical vapor deposition (PVD) [10], atomic layer deposition (ALD)[11,12], hydrothermal synthesis [13], sulfurization of molybdenum[14-16] or molybdenum oxide film[17] and chemical vapor deposition (CVD) [14]. CVD-based growth has the potential to be integrated with in-line fabrication at foundries making it the most favorable technique used for the synthesis of monolayered $MoS_2$ having large crystallite size and continuity over a large area.

To elucidate the growth mechanism of $MoS_2$ films using CVD, recently a number of experiments has been reported. One of the earlier attempts in 2012 used pre-deposited Mo film followed by CVD based sulphurization to achieve single and few-layered $MoS_2$ on Si/$SiO_2$ substrate [14]. In 2014 using atmospheric pressure CVD and precursors in the form of sulfur and $MoO_3$ powders, S.Wang et al. demonstrated that the local change of Mo:S precursors ratio govern the variation of the shapes of the crystals grown from triangular to hexagon [18]. Growth of $MoS_2$ films using $MoO_3$ as precursor shows triangular crystals while $MoCl_5$ based growth results into continuous growth [19]. Single crystalline $MoS_2$ flakes with size larger than 300 µm was grown by suppressing the nuclei density thorough adjusting the distance between sources and the substrate [20].Single step synthesis of monolayered large area $MoS_2$ films were grown on variety of substrates by using the naturally formed gap between $SiO_2$ coated wafer and the substrate acting as reactor cavity[21]. In another experiment, $MoO_2$ was used as a precursor and growth kinetics of $MoS_2$ was elucidated. They observed formation of two types of seeding centers: (a) Mo-oxysulfide ($MoO_xS_{2-y}$, yÓx) nanoparticles resulting into triangular growth for few layers $MoS_2$ and (b) atomic scale $MoS_2$ monolayer clusters to grow monolayer and irregular polygon shape [22]. Zhe et al. observed that the monolayered $MoS_2$ grown from irregular polygonal shaped cluster decorated with S-Mo and Mo-zz edges in a comparable ratio with dominant Mo-zz edges formation grows. Reactant concentration has been found to facilitate 2D-planar nucleation mechanism responsible for monolayer/bi-layer growth while, higher concentration leads to self-seeding nucleation mechanism responsible for few-layered / multi-layered growth [23]. A comparison of growth quality using three different precursors namely $MoO_3$, ammonium heptamolybdate (AHM), and tellurium (Te) assistance using $MoO_3$ for large continuous films using CVD shows better quality of large size growth using only $MoO_3$ powder and Sulfur powder with a particular ratio of approx. 30:1 [24].Growth at varying temperature shows that $MoS_2$ tend to grow laterally in triangular shapes a temperature below 730 °C, while growth above 730 °C results into longitudinal growth with hexagonal shapes [25]. Inspite of number of notable attempts, a standard recipe for the growth of continuous large crystallites of $MoS_2$ remains as a challenge [26,27]. A proper understanding of the growth mechanism is desirable for the CVD-based process to achieve large crystallite continuous monolayered film of $MoS_2$ having fewer grain boundaries which limits the mobility of the charge carriers [28]. During CVD based growth of $MoS_2$ number of parameters like nature of the precursor used [29,30] orientation of the substrate [31], gas flow rate [23], growth temperature & duration [25], and use of promoters[32] are expected to affect the crystallite size, phase, and shape.

In this work, $MoS_2$ grown at various experimental conditions has been systematically investigated using spectroscopic tools to establish the recipe for growth of continuous monolayered $MoS_2$ film having limited grain boundaries. The role of grain boundaries formation and triangular domains has been elucidated to achieve large area $MoS_2$ atomically thin films for electronic and opto-electronic applications.

## Experimental setup

For $MoS_2$ growth highly pure molybdenum trioxide $MoO_3$ (99.9 % sigma Aldrich) and sulfur powder (99.9% Sigma Aldrich) were used as precursors. Figure 1(a) shows the schematic of the CVD setup and parameters of $MoS_2$ growth are shown in Fig. 1(b)-(c). In CVD, the quartz tube of length 120 cm and diameter 4.5 cm was used. Temperature profile of both the precursors ($MoO_3$ and S powder) during complete cycle of growth has been shown in Fig. 1(b) and gas flow rate for the complete experimental duration. 285 nm $SiO_2$/Si wafer was used as a substrate for growth. The substrate was first cleaned sequentially using the iso-propyl alcohol, water and ethanol through the sonication bath process for 15 min each. After sonication, the substrate was washed with deionized water and then dried using the argon gas and put inside the hot oven at 100 ℃ for 10 min. The polished surface of the substrate faces downward to the precursor above the $MoO_3$ powder. Precursors and substrate were placed inside the CVD furnace as schematically shown in Fig. 1(a) separated at 19 cm. The growth was carried out at 750 °C for 5 mins using a single zone CVD furnace at atmospheric pressure. The outlet of the quartz tube was bubbled through a water bucket ensuring an equilibrium argon pressure inside the tube. During growth the amount of $MoO_3$ was varied between 10 mg to 30 mg referred as SET-I keeping gas flow rate constant 100 sccm and weight of sulfur 200 mg. In SET-II, the weight of the second precursor sulfur was varied in the range 100 mg to 500 mg keeping other conditions like weight of $MoO_3$ (15 mg) and gas flow rate (100 sccm) fixed. Finally, the carrier gas flow rate (Argon) was varied in the range 50 sccm to 300 sccm referred as SET-III, while keeping the other two factors $MoO_3$ and sulfur fixed at 15 mg and 200 mg respectively.

**Characterization**

The growth of $MoS_2$ flakes were confirmed initially using an optical microscope (Leica) with 10× (NA 1.3), 50× was used for optical micrograph. The detailed morphology of grown samples was recorded using a field emission electron microscope (ZEISS Sigma 300 FESEM). The number of layers of $MoS_2$ was characterized using a Raman spectrometer (Renishaw Invia-Raman Spectrometer) in the range 350-450 nm. The photoluminescence spectra of samples were recorded in the range 500-800 nm using $\lambda_{ex}$= 514 nm (laser excitation).

**Result and discussion**

Figure 2(a) shows the optical microscope image of the grown continuous film of $MoS_2$ sample. FESEM image of individual triangle at higher magnification grown continuous sheet of $MoS_2$ film is shown in Fig. 2 (b). Raman spectra shows two vibration modes at 385.7 ($E^1_{2g}$) and 405 $cm^{-1}$($A_{1g}$) as shown in figure 2(c). Observed frequency difference of 19.5 $cm^{-1}$ confirms growth of monolayer $MoS_2$ in agreement with previous work [14,24] . The frequency difference between these two modes depends upon the number of layers [33]. Both the in plane vibrational mode ($E^1_{2g}$) and out of plane ($A_{1g}$) modes shows requirement of two-peak structure for fitting the experimental curve as shown in Fig.2 (c). Fitted line profile is summarized in Table-I. Interestingly, the in-plane vibrational mode ($E^1_{2g}$) mode is Gaussian while out of plane ($A_{1g}$) mode is Lorentzian in line-shape. Such two peak structure / splitting of in plane vibrational mode ($E^1_{2g}$) was previously observed in $MoS_2$ with lattice strain of more than 1 %, which arises due to disorder in the crystalline structure of the material [34].The photoluminescence spectra recorded from same position is shown in Fig. 2(d). The experimental curve was fitted using Gaussian line shape which shows emission peak at 679.2 nm (1.83 eV) with shoulders at 624.4 nm (1.99 eV) with relative intensities of 4421 counts and 1869 counts respectively. The 1.83 eV peak is assigned to the A-exciton while 1.99 eV peak arises due to B-excitons [35,36]. A-exciton shows peak width $\Delta\omega$ = 44.1 nm, while the B-exciton shows $\Delta\omega$ = 63.7 nm. Observed strong A-exciton peak relative to the B-exciton indicates monolayer growth [5] in agreement with the Raman spectra shown previously. R. Coehoorn et al. observed two prominent peaks at 670 and 627 nm in the absorption spectrum. These two resonances have been established to be the direct excitonic transitions at the Brillouin zone K point. Their energy difference is due to the spin-orbital splitting of the valence band. [35,36].

Fig. 3 shows the effect of $MoO_3$ weight fraction during growth. Fig. 3(a) shows nucleation of small equilateral triangular domains randomly on the substrate having crystallite size ~7 μm when 10 mg of $MoO_3$ was used along with 200 mg of sulfur. Lower concentration of $MoO_3$ results in a lower growth rate. Optical image of growth with 15 mg $MoO_3$ indicates formation of continuous films with large crystal size ~ 30 μm . Increased $MoO_{3-x}$ vapor flux with increasing weight fraction of $MoO_3$ leads to increased growth rate resulting into formation of continuous films with grain boundaries as visible in Fig. 3(b). Grain boundaries of continuous film formation were observed due to low mass flux and high growth rate [37],Coalescence of the particles takes place resulting into formation of continuous sheet with increased $MoO_3$ weight fraction to minimize the surface energy through reduction of surface area. With further increase of $MoO_3$ concentration to 20 mg, multi-nucleation sites were observed under optical microscope as shown in Fig. 3(c). In this case we observe continuous film with discrete crystal size range from ~7 to 25 μm (Fig. 3(c)). For 25 mg of $MoO_3$, the crystal shape changes from sharp edged triangles to triangles with rounded corners having crystallite size of 7 μm as shown in Fig 3(d). Further increase of $MoO_3$ weight fraction to 30 mg results into triangular crystals with smaller size ~7 μm and higher nucleation density on the substrates as shown Fig. 3(e). Growth with increasing weight of $MoO_3$ relative to sulfur results into large crystal of $MoS_2$ flakes with continuous films tuned by Vapor phase of $MoO_{3-x}$. The precursor evaporates in the form of flux, resulting into different size nanostructures. The deposition of $MoO_{3-x}$ molecular clusters are reduced to $MoS_{2-x}$ clusters (bright spots inside the triangle) in the initial stage of the growth process, consequently leading to a 2-dimensional nucleation on $SiO_2$/Si substrate (as shown schematically in figure S1 – Supplementary information)[38]. The growth of $MoS_2$ is limited by the diffusion of vapor-phase $MoO_{3-x}$ [39]. Tuning of weight fraction of precursor also affects Mo:S molar ratio along the substrate which affects the morphology of $MoS_2$, on the basic principle of crystal growth [18]. As we increase the amount of $MoO_3$ to 15 mg, the density of $MoO_{3-x}$ clusters increases gradually with a proper ratio 1:60 of Mo:S atom concentration which forms a continuous film of $MoS_2$ flakes. Higher concentration of $MoO_3$ leads to multiple $MoS_{2-x}$ clusters resulting into formation of multilayer growth. 30 mg $MoO_3$ precursor leads to formation of multiple bright spots due to incomplete sulfurization of number of clusters [22,23]. With increasing $MoO_3$ weight fraction the coverage area monotonically increases while the variation in domain size shows a maximum at 15 mg of $MoO_3$ as shown graphically in fig. 3(f). (Figure.S1- Supplementary information) schematically shows the variation of the nucleation centers density with increasing $MoO_3$ concentration. Interestingly, even in the past experiments, 15 mg of $MoO_3$ has been found to result into continuous sheet of $MoS_2$ [18,21] and the concentration of Mo governs the shape and size of Monolayer $MoS_2$ crystals [40].

To understand the change in the crystal structure and to quantify the number of layers, samples of SET-1 were subjected to Raman spectroscopy measurements. Spectra were collected from different regions of the sample as shown in Fig. 4(a). Raman spectra shows two prominent peaks at ~ 384 $cm^{-1}$ and ~ 406 $cm^{-1}$ assigned to in-plane $E^1_{2g}$ vibration mode and out of plane $A_{1g}$ vibration mode. In plane $E^1_{2g}$ vibration mode arises due to in plane vibration of Mo and S atoms, while out of plane $A_{1g}$ vibration mode arises due to out of plane vibration of Sulphur atoms [33]. The peak width of these modes was estimated upon fitting $E^1_{2g}$ and $A_{1g}$ modes with Gaussian and Lorentzian line-shape respectively as summarized in Table-I (fitted curves shown in Supplementary information Figure S2). Lorentzian line shape indicates high crystallinity of as grown $MoS_2$ films. The difference in the peak position of $E^{1}_{2g}$ and $A_{1g}$ modes indicates the number of layers due to weak van der waals interlayer coupling, and the presence of coulombic interlayer interaction in $MoS_2$ [33] . $E^{1}_{2g}$ and $A_{1g}$ modes are shifted as additional layers added to form the bulk material from individual layer because interlayer van-der wall interaction increase the effective restoring force acting on the atoms [41].With decreasing number of layers $A_{1g}$ mode shifts closer to $E^{1}_{2g}$ modes [33]. Sample grown with 10, 20, 25 and 30 mg shows peak separation of 25.2 $cm^{-1}$, 23.8 $cm^{-1}$, 19.8 $cm^{-1}$ and 22.6 $cm^{-1}$. Raman spectra of samples grown with 25 mg of $MoO_3$ indicated monolayered growth. While growth with 30 mg of $MoO_3$ indicates bilayer growth as evident. The higher relative concentration of $MoO_3$ as compared to sulfur leads to initialization of multi-nucleation sites resulting into multilayer growth, while growth with 25 mg of $MoO_3$ results into monolayered growth. In the case of layered growth the deposit $MoO_3$ ñwetsò the substrate due to dominance of surface tension between substrate-vapor ($\gamma_{sv}$) over combined effect of surface tension between film-substrate ($\gamma_{fs}$) and film-vapor ($\gamma_{fv}$) ie. $\gamma_{sv} \geq \gamma_{fs} + \gamma_{fv}$ . [38] Fig. 4(b) shows the photoluminescence spectra of samples. The spectral line profile was obtained by fitting with the gaussian lineshape as summarized in Table-II (fitted curves shown in Supplementary information Figure S3). Grown sample with 10 mg of $MoO_3$ shows peaks at ~ 671 nm (1.85 eV) and 628.6 nm (1.97 eV) corresponding to A-exciton and B-exciton respectively. Fitting of PL spectra requires two Gaussian peaks for A-exciton in addition to B-peak at ~ 671.4 nm (1.85 eV) and 681.3 nm (1.82 eV) respectively. A and B exciton arises due to spin-orbit coupling induced splitting of valence band [5]. $MoO_3$-10 shows a relatively weaker luminescence peak, indicating multilayer growth in agreement with the observed Raman spectra. With increase of the $MoO_3$ weight fraction to 20 mg leads to observation of highest PL intensity and with further increase the PL intensity monotonically decreases. Strong emission indicates presence of direct band, high crystalline quality of grown monolayer films. With increasing number of layers to bi-layers and tri-layers, the luminescence intensity gets suppressed and disappears [5]. At lower concentration (10 mg) and moderate 15 mg concentration of $MoO_3$, we observed a planar flake but with higher concentration (20 and 25 mg) the flakes contended nanocrystals on the $MoS_2$ flakes. At much higher concentration (30 mg) the size decreased with increase in the density of nanocrystals over the $MoS_2$ flakes**.** Thus, concentration of $MoO_3$ is responsible for self-seeding process and their concentration decide the nucleation site and coverage area. At much higher concentration the density of nanocrystals increases indicating incomplete sulphurization. Similar observations has been reported by D. Zhu et al. [22]. Samples grown with 25 mg and 30 mg of $MoO_3$ shows two peak structures for the direct-gap transition referred as A-exciton peak (i.e A-exciton peak and $A^-$- trions peak ) [42]. Previously negative doping of $MoS_2$ by $SiO_2$ substrate has been accounted for the origin of A-trion emission and observed red shift with broadened main peak in PL [43]. Interestingly in our experiment, except weight fraction of $MoO_3$ all other parameters were kept fixed, observation of A- trion in samples grown using higher $MoO_3$ weight fractions exclusively contradicts the hypothesis proposed previously in terms of substrate induced negative doping effect, since the common substrate was used in all sets of experiments. In another work, observation of $A^-$-trions peak has been accounted to high purity interface between the sample and the substrate [20].

Fig. 5 shows growth of $MoS_2$ with varying amount of sulfur relative to $MoO_3$ (SET-II), keeping weight of $MoO_3$ powder fixed at 15 mg and gas flow rate 100 sccm. Fig. 5(a) shows results with 100 mg of sulfur. $MoS_2$ flakes at random locations on the substrate were formed having shapes varying from triangle to star like pattern with crystallite size of ~11 [illegible]. On substrate, there is a small triangular growth of $MoS_2$ flakes due to lower concentration of sulfur in the growth atmosphere near the surface. Doubling the sulfur amount to 200 mg results into crystallite size of ~ 30 μm and continuous film upto few $mm^2$ as observed in Fig. 5(b). With further increase of weight fraction to 300 mg of sulfur, $MoS_2$ crystals with triangular shape and varying crystallite sizes ~ 7 to 25 μm were observed, Fig. 5(c). Also crystallite domains were not continuous as observed at lower weight fractions. If the weight of the sulfur is increased to 400 mg, the crystallite size of the triangular $MoS_2$ flakes is reduced to ~ 7 μm due to decrease in lower concentration of Mo:S atom ratio, Fig. 5(d). With excessively large weight fraction of sulfur to 500 mg, formation of star like flakes were observed having average crystallite size of about ~ 20 μm (Fig. 5(e)). A closer observation to the center of the star flakes indicates the presence of clusters responsible for multinucleation. Due to the sufficient amount of sulfur, multi-layer formation is observed. Proper weight fraction of sulfur has been observed as the prime factor deciding formation of number of layers/ crystallite size. It's clear from the optical micrographs that a proper amount of Mo:S ratio of the precursor plays critical role towards maintaining triangular shape [44]. With varying concentration of Sulfur, the nucleation density significantly changes in contrast to previous report where the role of sulfur towards nucleation was found to be negligible [20].

Fig. 6 shows optical micrograph of $MoS_2$ samples grown with varying gas flow rates (50 to 300 sccm) keeping precursors weight $MoO_3$ (15 mg) and Sulfur (200 mg) constant. Growth with 50 sccm of Argon flow results into $MoS_2$ flakes having large crystallite size of ~ 80 μm and coverage area of ~ 40 % as shown in Fig. 6(a). With moderate gas flow rates (100 and 150 sccm) the discrete crystals of $MoS_2$ flakes join together to form continuous films. Gas flow rate of 100 sccm results into coverage area of 70 %, while flow rate of 150 sccm leads to 92 % coverage area as shown in figure 6(b). Further increase of gas flow rate to 200, 250 and 300 sccm results into smaller crystallites with discrete regions. A gas flow rate of 300 sccm results into $MoS_2$ crystallites of size as small as ~ 7 ⍰⍰ . It is obvious from the optical micrographs shown in Fig. 6 (c)(d)and(e) respectively that carrier gas flow rate is another key parameter which influence the crystallite size and coverage area by tuning the local concentration of $MoO_{3-x}$ and sulfur vapor precursors being nucleated on the substrate. Carrier gas flow rate decides the residence time of vapors of precursors on the substrate deciding nucleation density, crystallite size and coverage area. Under sufficient supply of precursor and nucleation sites, the growth extends in the form of large-area single-layered $MoS_2$ continuous film with increasing gas flow rate as shown in figure 6(f). It is also interesting to note that with increase of the crystallite size, the coverage area decreases.

Fig. 7 shows effect of change of molar ratio of Mo and Sulfur on the crystallite size and on the coverage area. The change in the morphology of $MoS_2$ along the gas flow direction in the Mo: S atom ratio along the Si substrate surface. With increasing molar ratio from 30:1 to 60:1 the crystallite size increases. With further increase of the molar ratio, the crystallite size monotonically decreases for the molar ratios 90:1 and 120:1. Growth using increased molar ratio to 150:1 shows increase in the crystallite size. A molar ratio of 60:1 results into $MoS_2$ crystals of average size ~30 μm, while molar ratio of 60:1 results into continuous films. $MoS_2$ crystals grown with molar ratio of 90:1 shows the strongest photoluminescence intensity, indicating monolayer growth. Under all attempted growth conditions by varying precursorôs weight fractions, gas flow rates, we obtained triangular shaped $MoS_2$ growth. While V Senithikumar et al. [45] observed variation in the grown crystallites due to growth under low pressure. Itôs interesting to observe that the crystallite shape remains triangular due to growth under atmospheric pressure conditions using single zone furnace without any requirement of vacuum conditions.

## Conclusion

Through controlled growth at various experimental parameters, we have been able to successfully grow continuous films of $MoS_2$ monolayers on $SiO_2$/Si substrate through CVD. Optimized precursor weight ratio i.e $MoO_3$ and Sulphur (15 mg: 200 mg) results into discrete crystallite size as large as 80 μm with 50 sccm flow of Argon as carrier gas and coverage area of ~ 40 %. Thus, concentration of $MoO_3$ is responsible for self-seeding process and their concentration decides the nucleation site and coverage area. With moderate gas flow rates (100 and 150 sccm) the discrete crystals of $MoS_2$ flakes join together to form continuous films. Gas flow rate of 100 sccm results into continuous film with 70 % coverage area. While flow rate of 150 sccm leads to 92 % coverage area. Increase of the crystallite size leads to decrease in the coverage area. $MoO_{3-x}$ was found to be responsible for self-seeding nucleation and its amount decides the nucleation density. The formation of $MoS_2$ flakes and its shape is dependent on local concentration which is tuned by the relative amount of $MoO_3$ and sulfur precursors. The carrier gas flux decides the residence time of Mo and S on the substrate towards growth. The carrier gas flow rate is responsible for local concentration of vapor precursors and their coverage area.

## Acknowledgements

One of the authors Rakesh K. Prasad thanks TEQIP-III for fellowship. Dilip K. Singh thanks DST, Government of India (Fellowship IFA13-PH65; CRG/2021/002179; CRG/2021/003705) and Seed Money Scheme, Birla Institute of Technology, Mesra for funding.

## Supplementary information

Supplementary information contains schematic diagram of effect of concentration of $MoO_3$ precursor on self seeding sites along with fitted Raman and photoluminescence spectra plots for $MoS_2$ flakes grown under different concentration of $MoO_3$.

## Data availability statement

The datasets generated during and/or analyzed during the current study are available from the corresponding author on reasonable request.

**Table I:** Fitted parameters of Raman spectra

| Sample | *Lineshape Parameters* | Peak-1 | Peak-2 | Peak-3 *Gaussian* | Peak-4 | Peak-5 | Peak-6 *Lorentzian* |
|---|---|---|---|---|---|---|---|
| $MoO_3$-10 | *Center* | | | 380.2 | 383.7 | | 408.9 |
| | *FWHM* | - | - | 10.1 | 4.5 | - | 5.2 |
| | *Ampl.* | | | 1044 | 2299 | | 6424 |
| $MoO_3$-15 | *Center* | | | 382.1 | 385.7 | 405.2 | 408.9 |
| | *FWHM* | - | - | 9.9 | 3.1 | 6.5 | 5.2 |
| | *Ampl.* | | | 1118 | 7125 | 7280 | 6424 |
| $MoO_3$-20 | *Center* | 342.3 | 365.8 | 380.9 | 383.8 | 407.6 | 409.2 |
| | *FWHM* | 34.1 | 17.0 | 10.6 | 4.5 | 6.7 | 4.2 |
| | *Ampl.* | 773 | 649 | 1984 | 3540 | 9641 | 4301 |
| $MoO_3$-25 | *Center* | 349.3 | 366.1 | 382.4 | 384.8 | 404.6 | |
| | *FWHM* | 8.3 | 10.2 | 10.0 | 3.4 | 5.8 | - |
| | *Ampl.* | 220 | 450 | 865 | 4332 | 2171 | |
| $MoO_3$-30 | *Center* | | | 382.6 | 384.1 | 406.7 | 408.9 |
| | *FWHM* | - | - | 11.1 | 3.0 | 5.7 | 3.4 |
| | *Ampl.* | | | 2866 | 16631 | 10839 | 11872 |
| | *Assignment (Lineshape)* | | | $E_{2g}$ *Gaussian* | $E_{2g}$ *Gaussian* | $A_{1g}$ *Lorentzian* | $A_{1g}$ *Lorentzian* |

**Table II:** Fitted parameters of PL spectra using Gaussian line-shape

| Sample | *parameters* | Peak-1 | Peak-2 | Peak-3 | Peak-4 | Peak-5 | B/A ratio |
|---|---|---|---|---|---|---|---|
| $MoO_3$-10 | *Center* | 628.6 | 671.4 | - | 681.3 | - | |
| | *FWHM* | 39.2 | 21.5 | | 48.2 | | |
| | *Ampl.* | 63 | 263 | | 244 | | |
| $MoO_3$-15 | *Center* | 624.4 | - | 679.2 | - | - | |
| | *FWHM* | 63.7 | | 44.1 | | | |
| | *Amp* | 1869 | | 4421 | | | |
| $MoO_3$-20 | *Center* | - | 669.8 | - | 687.5 | 695.5 | |
| | *FWHM* | | 15.77 | | 49.4 | 132.4 | |
| | *Ampl.* | | 1697 | | 7994 | 2044 | |
| $MoO_3$-25 | *Center* | 627.7 | 670.6 | 673.5 | 679.2 | - | |
| | *FWHM* | 38.6 | 17.1 | 148.9 | 42.7 | | |
| | *Ampl.* | 722 | 2796 | 860 | 3651 | | |
| $MoO_3$-30 | *Center* | 627.8 | 663.9 | 661.8 | 677.5 | | |
| | *FWHM* | 44.7 | 16.5 | 143.2 | 43.03 | | |
| | *Ampl.* | 911 | 1294 | 1096 | 3425 | | |
| | *Assignment* | *B-exciton* | *A-exciton* | *$A^-$-trion* | | | |

**Figure captions:**

Figure-1 (a) Schematic setup for thermal CVD for $MoS_2$ synthesis (b) Temperature variation for $MoO_3$ and Sulfur during $MoS_2$ growth Process. (c) Variation of gas flow rate during growth.

Figure-2 (a) Optical Microscope image of grown $MoS_2$ flakes (b) FESEM images of $MoS_2$ grown on 285 nm - $SiO_2$ /Si (c) Raman Spectra of $MoS_2$ (d) Photoluminescence spectra from grown $MoS_2$ films.

Figure-3 Optical micrographs of CVDï grown $MoS_2$ flakes with varying amounts of $MoO_3$ precursor (a) 10 mg (b) 15 mg (c) 20 mg (d) 25 mg (e) 30 mg (f) Variation of domain size and coverage area with increasing weight of $MoO_3$.

Figure-4 (a) Raman spectra of grown $MoS_2$ with varying $MoO_3$ weight fraction (b) Photoluminescence spectra.

Figure-5 Optical Microscope images for the CVD-grown $MoS_2$ flakes with varying amounts of sulfur (a) 100 mg (b) 200 mg (c) 300 mg (d) 400 mg (e) 500 mg, and (f) Dependence of domain size and coverage area on the weight fraction of the sulfur.

Figure-6 Shows the optical image for CVD grown $MoS_2$ flakes with varying gas flow rate (a) 50 sccm (b) 150 sccm (c) 200 sccm (d) 250 sccm (e) 300 sccm (f) shows dependence of coverage area and crystallite size on carrier gas flow rate.

Figure-7 Effect of molecular ratio ($MoO_3$: Sulfur) on the (a) Crystallite size and (b) Coverage area %.

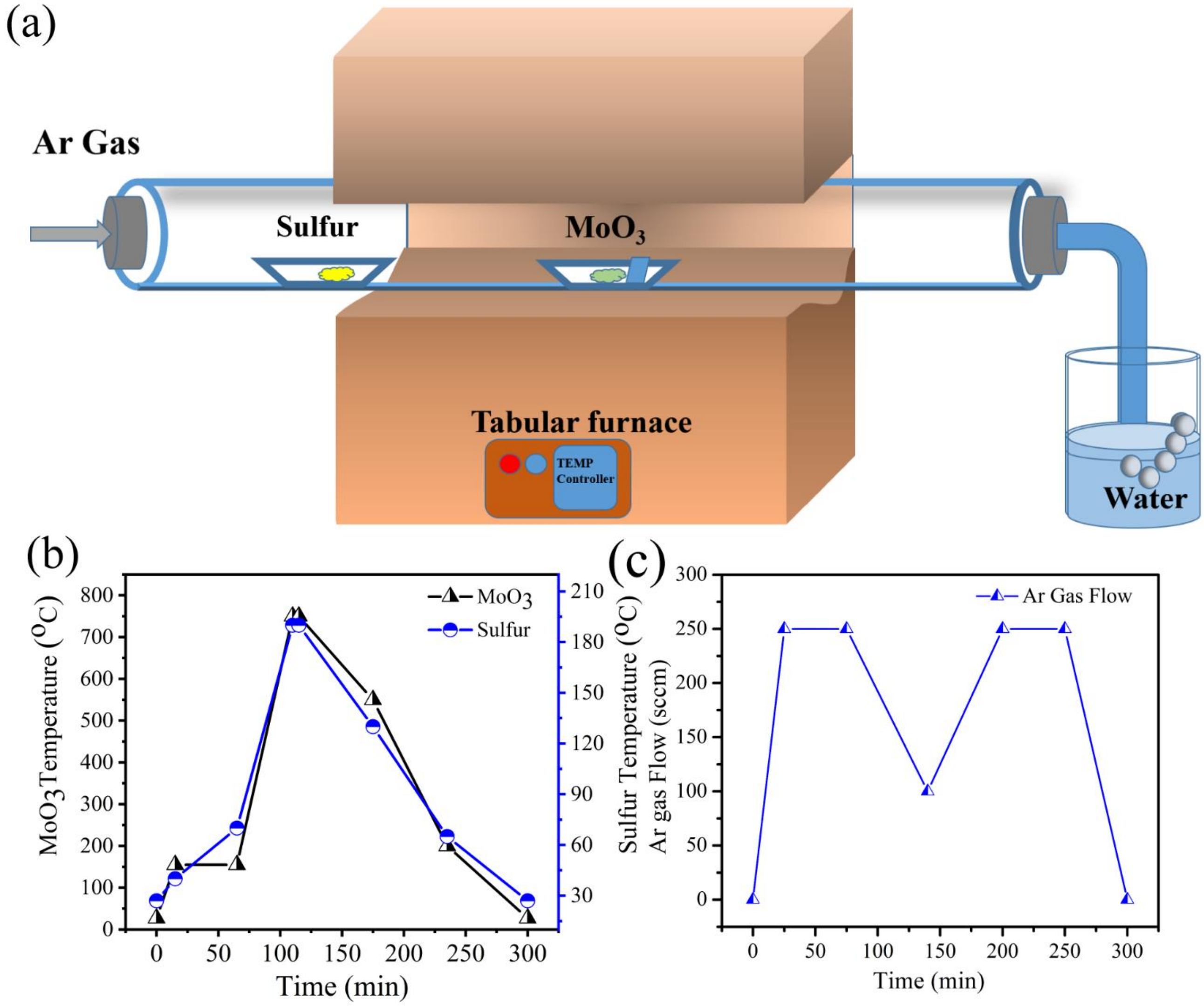

**Fig. 1 Singh et al.**

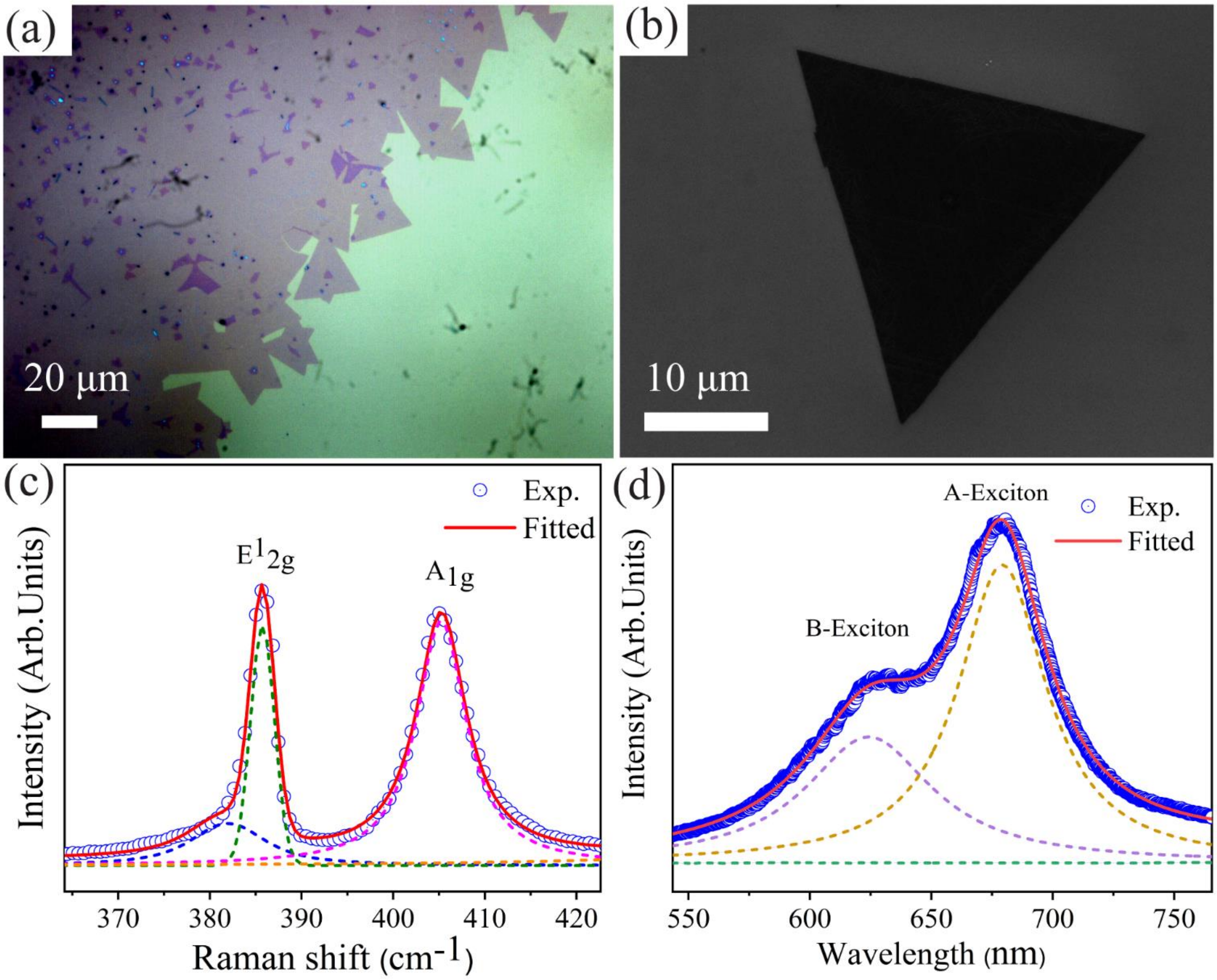

**Fig. 2 Singh et al.**

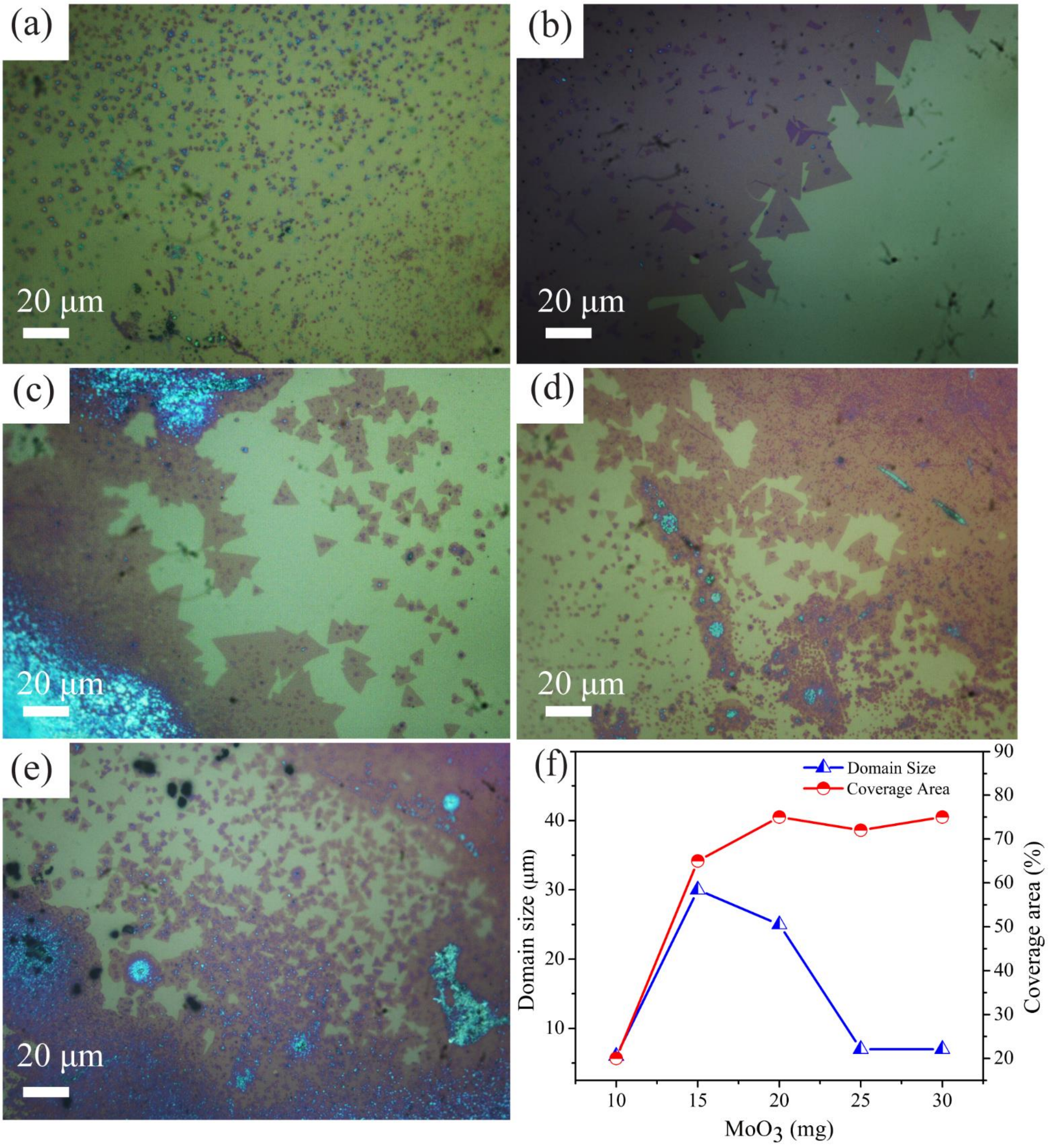

**Fig. 3 Singh et al.**

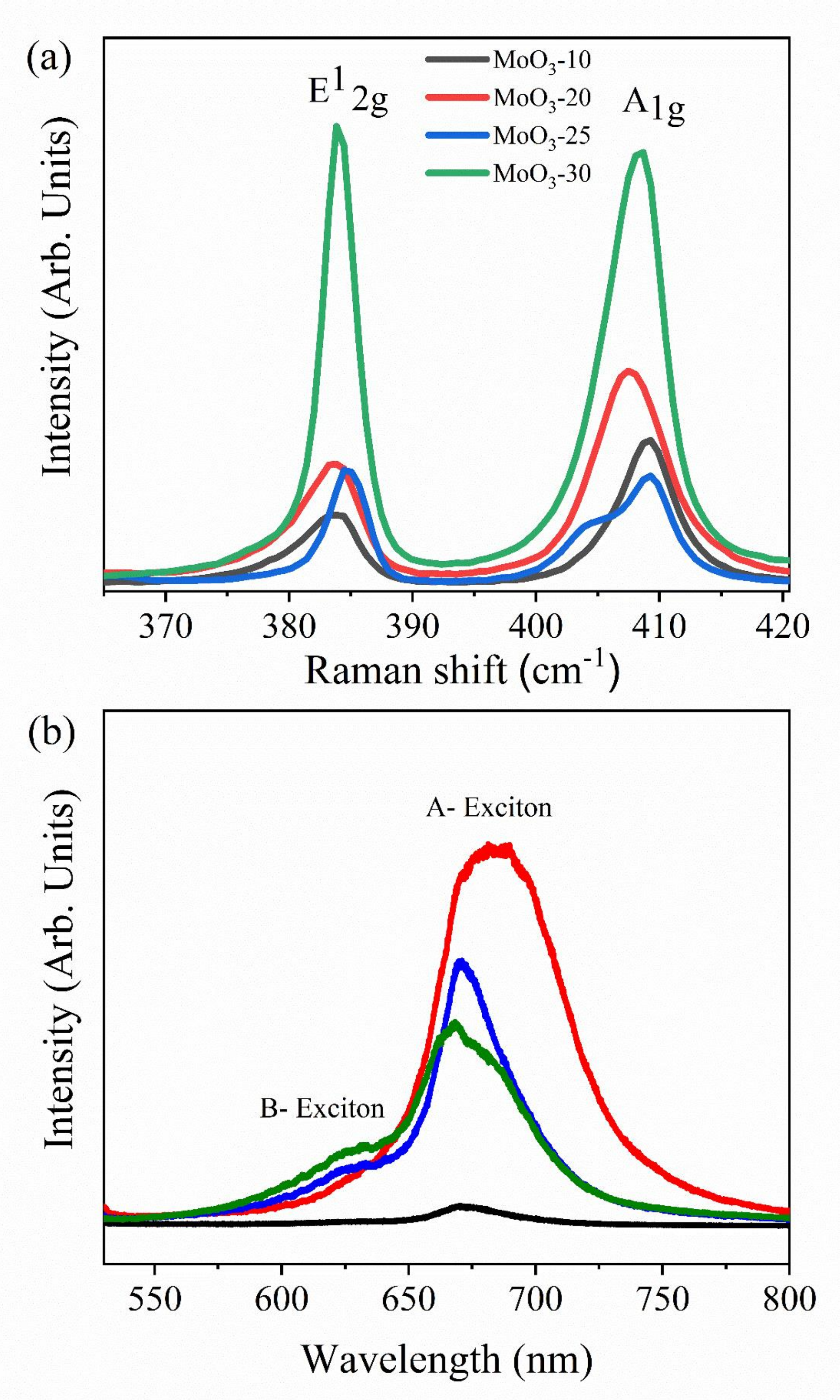

**Fig. 4 Singh et al.**

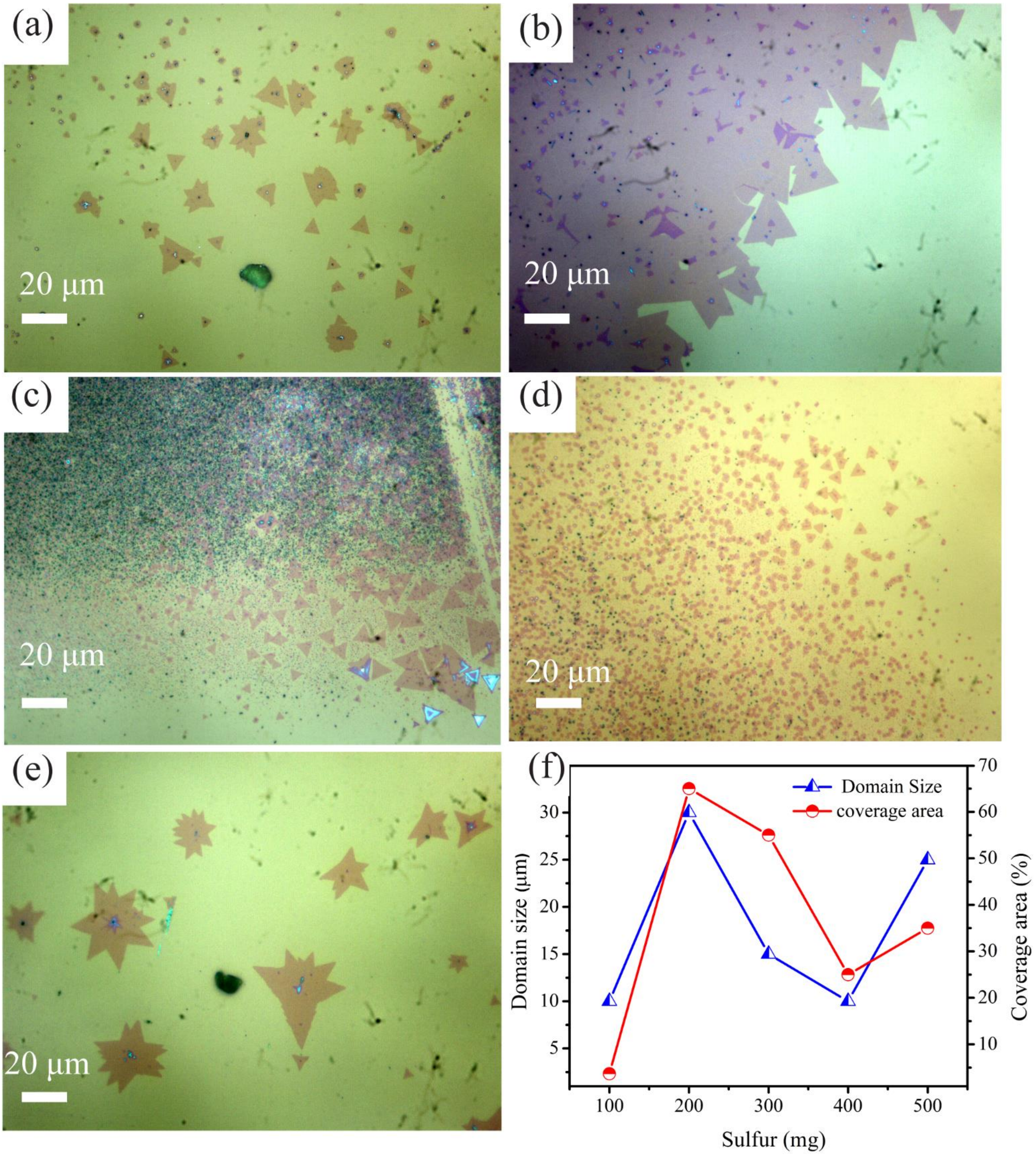

**Fig. 5 Singh et al.**

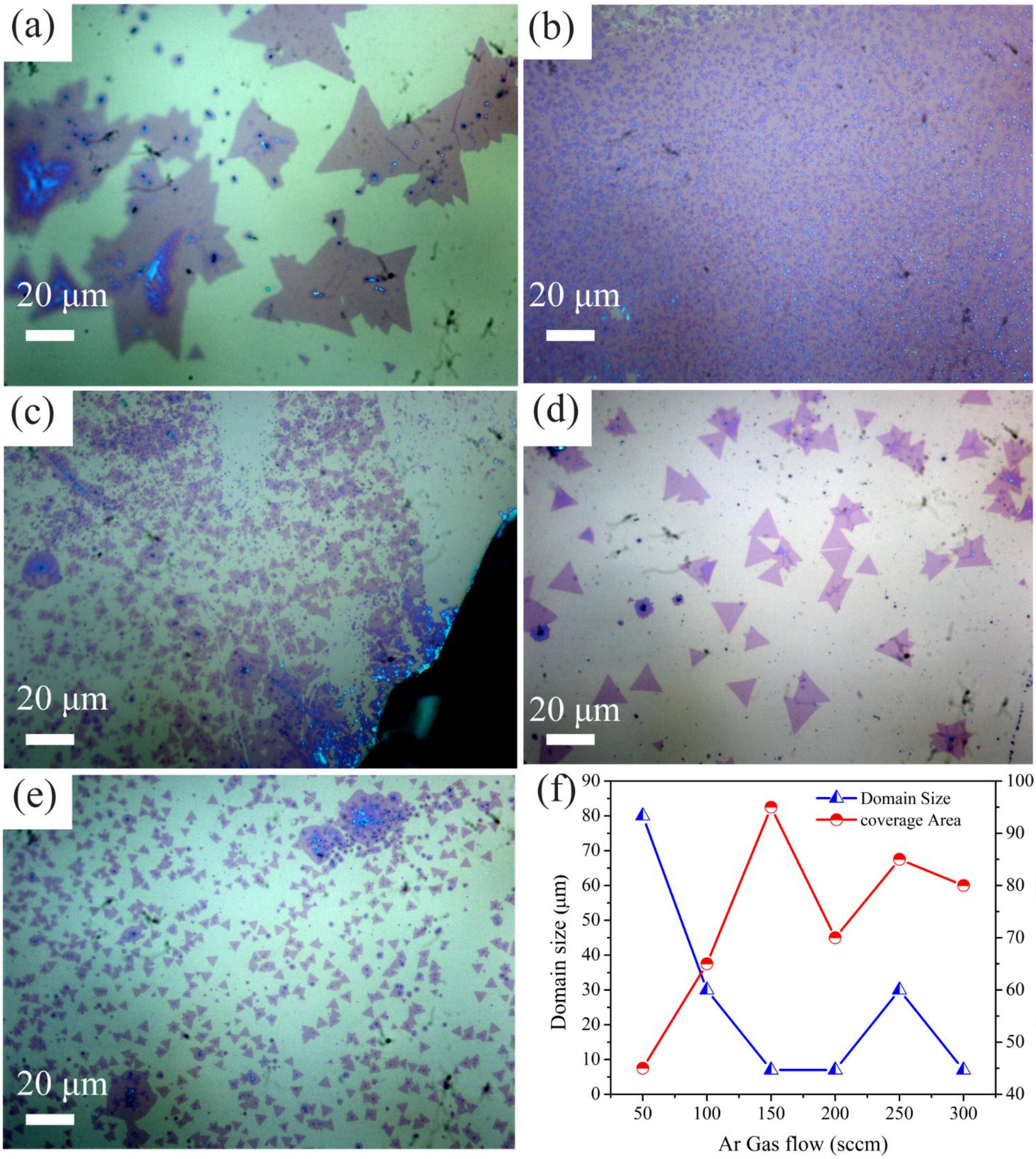

**Fig. 6 Singh et al.**

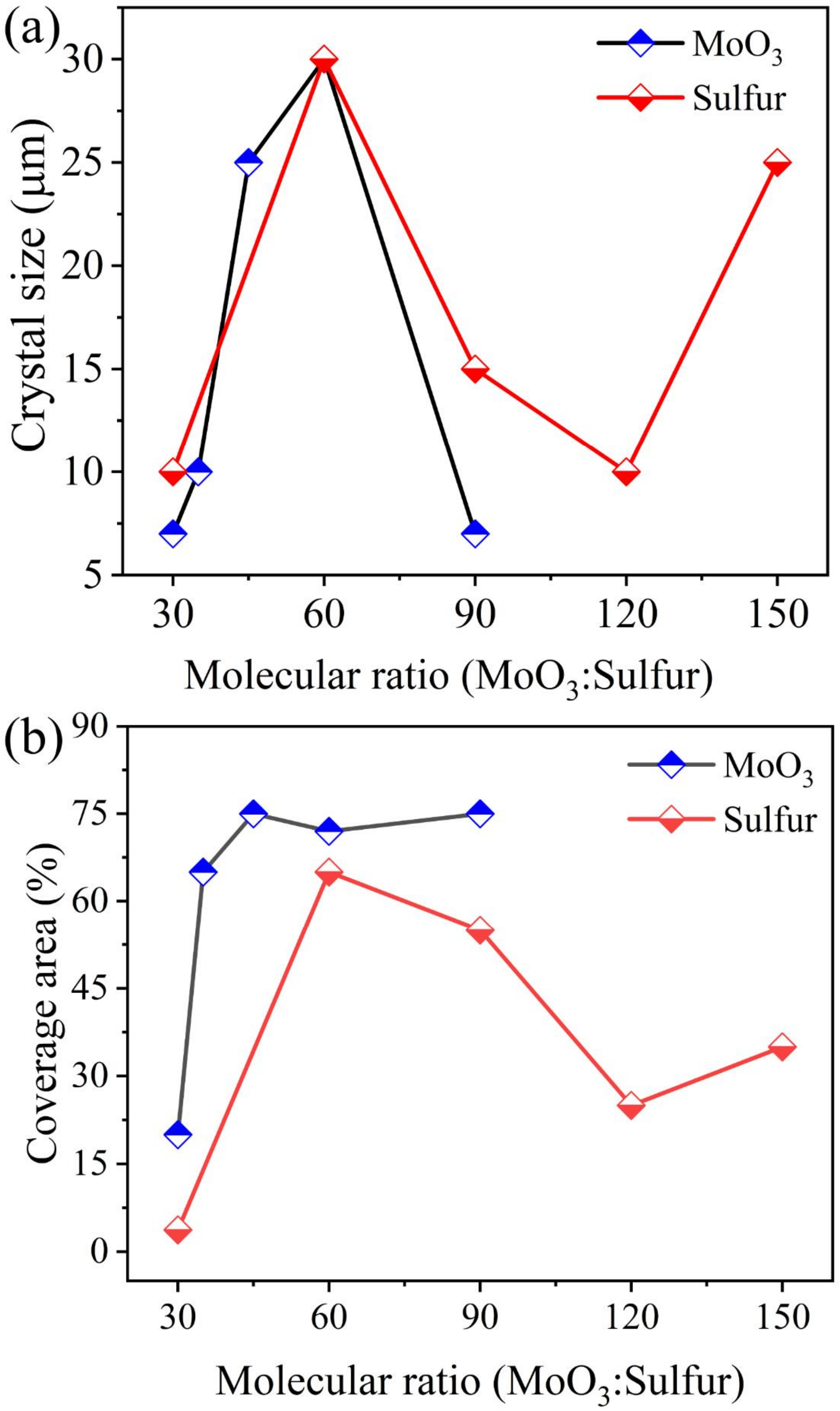

**Fig. 7 Singh et al.**

**Supplementary information**

# Kinetics and Parameters of Epitaxial Monolayered Continuous large area Molybdenum disulfide Growth

Rakesh K. Prasad[1], Dilip K. Singh[1*]

[1]Birla Institute of Technology Mesra, Ranchi -835215 (India)

[*]Email: dilipsinghnano1@gmail.com

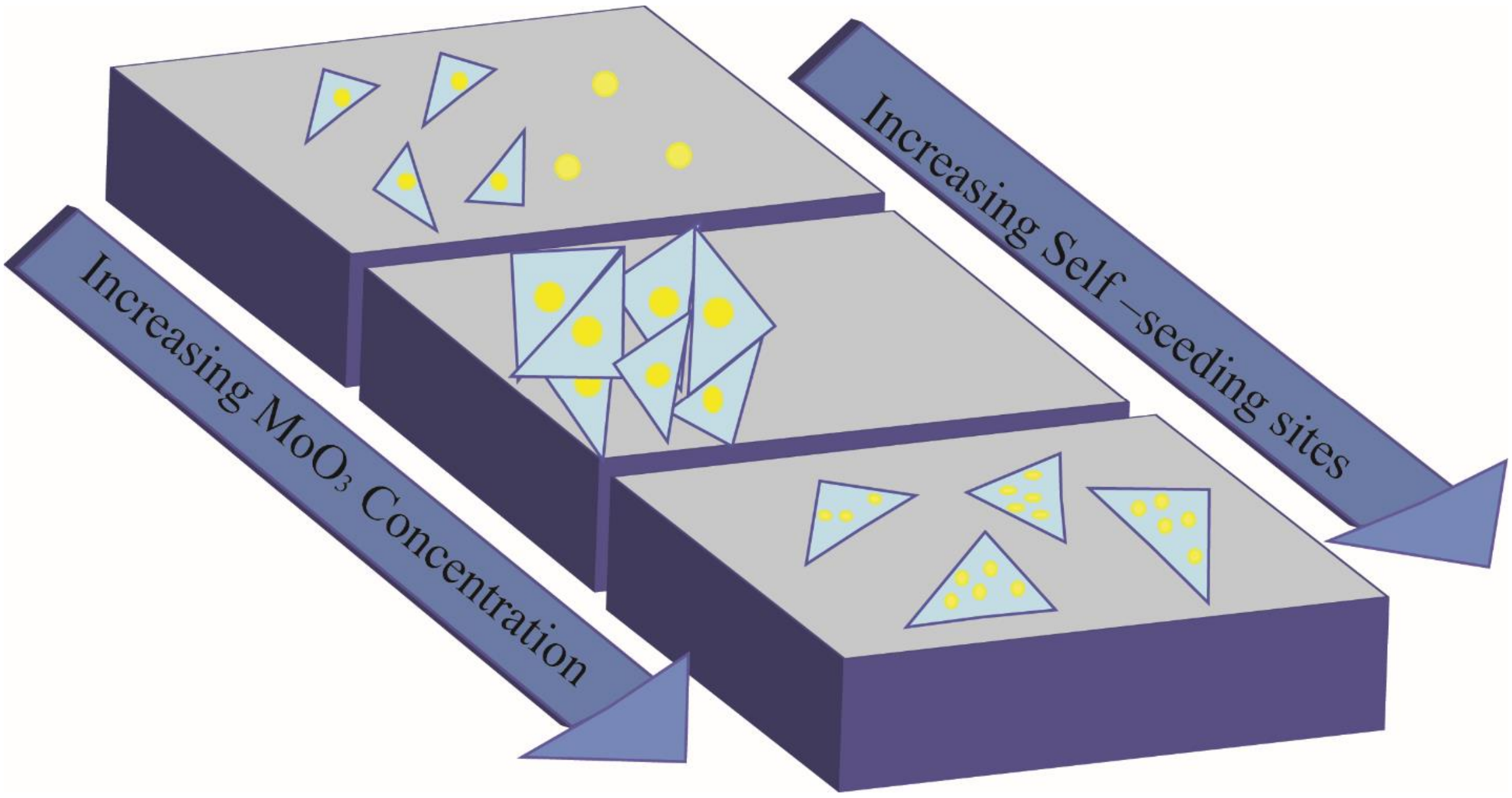

**Fig. S1**: Schematic diagram, illustrate the effect of concentration of $MoO_3$ precursor on self - seeding sites

Figure S1 schematically shows that lower concentration of $MoO_3$ helps to grown smaller crystal size of $MoS_2$ flakes with single seeding sites and with moderate amount of $MoO_3$. During growth, keeping other parameter constant like sulfur, gas flow rate and position of substrate helps to grow large crystal size. Higher amount of $MoO_3$ concentration increases the density of self-seeding sites inside the single flakes which is further responsible form multi-layer growth.

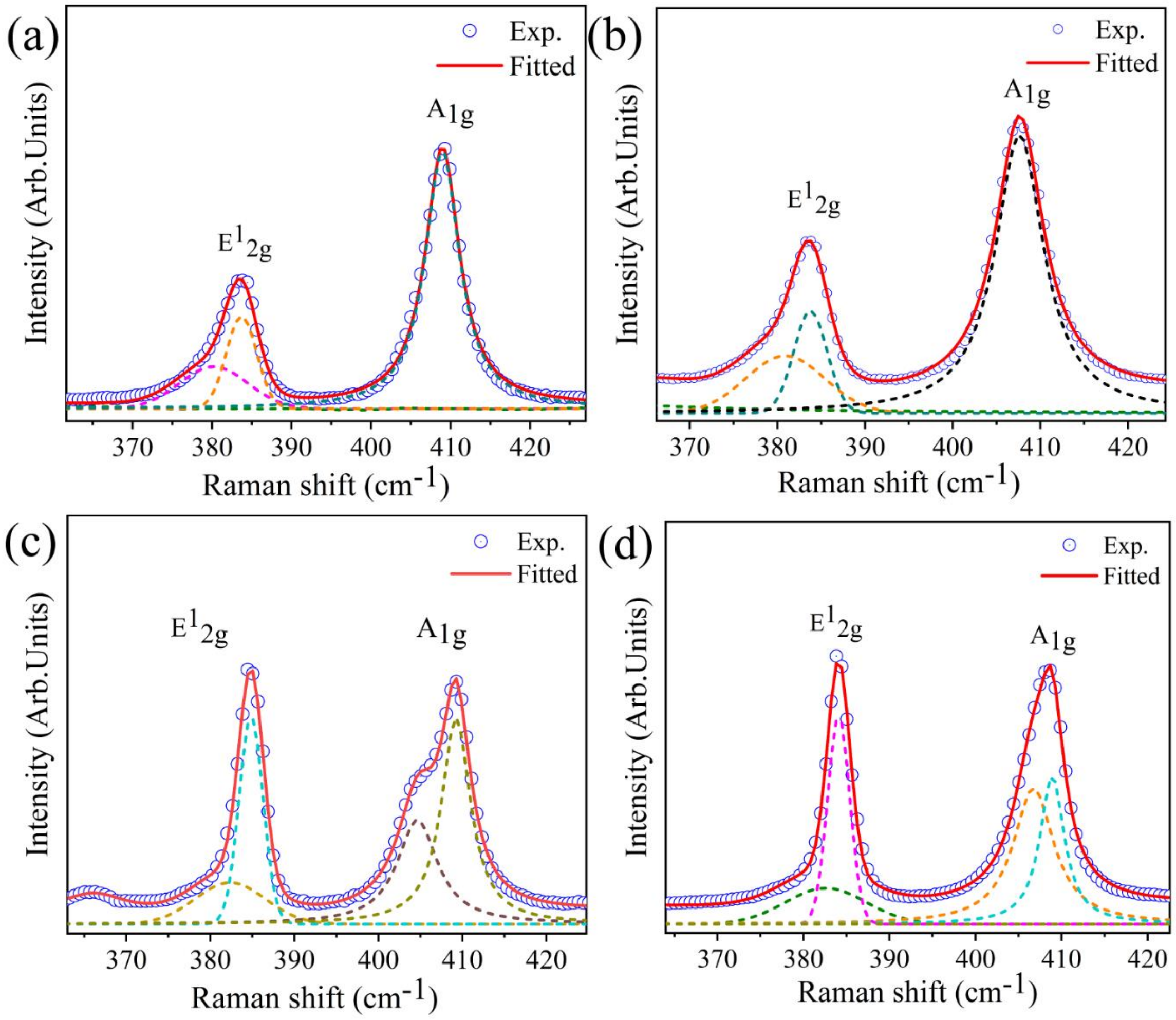

**Fig. S2**: Raman spectra plot for $MoS_2$ flakes grown under different concentration of $MoO_3$ (a) 10 mg (b) 20 mg (c) 25 mg (d) 30 mg

Fig. S2 shows fitted Raman specta of grown $MoS_2$ sample with varying concentration of $MoO_3$ keeping other parameters constant. With lower concentration of $MoO_3$ (10 mg and 20 mg), $A_{1g}$ shows single peak, but with higher concentration of $MoO_3$ (25 mg and 30 mg), $A_{1g}$ peak requires two peaks to fit indicating incomplete formation of multilayer on monolayer $MoS_2$.

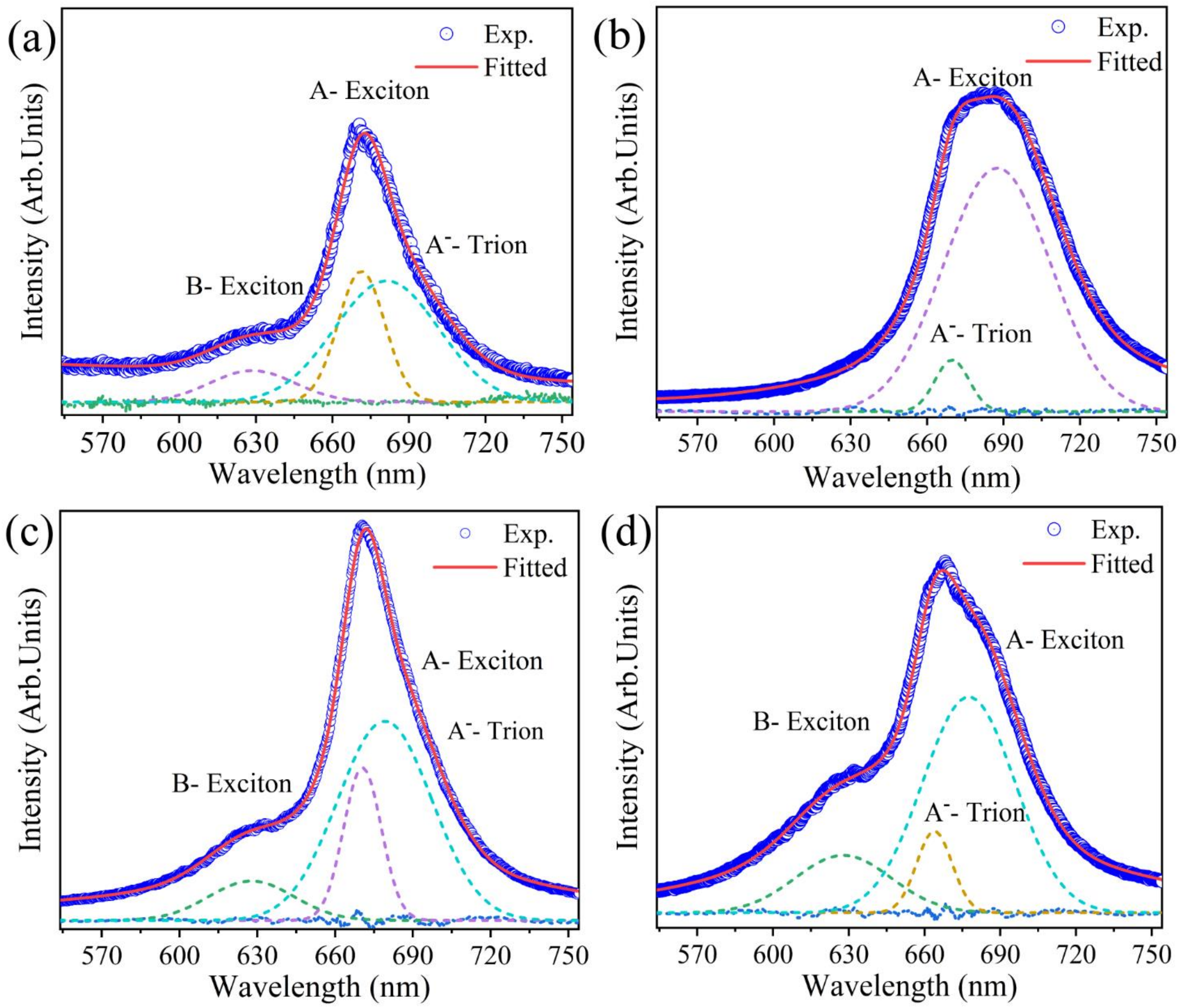

**Fig. S3:** Photoluminescence spectra plot for $MoS_2$ flakes grown under different concentration of $MoO_3$ (a) 10 mg (b) 20 mg (c) 25 mg (d) 30 mg

Fig. S3 show fitted photoluminescence spectra of $MoS_2$ grown by CVD using varying concentration of $MoO_3$. Three spectra line (A ï Exciton, $A^-$-Trion and B- Exciton) are observed at ~ 627,~ 670, and ~ 680 nm. 627 nm peak indicates B-Exciton 670 nm peak indicates A-Exciton and 680 nm peak indicates $A^-$ -Trion, All peaks indicate high crystal quality of film growth of $MoS_2$ crystals.